\documentclass[superscriptaddress,twocolumn]{revtex4-1}
\pdfoutput=1
\usepackage[utf8x]{inputenc}
\usepackage{lineno}
\usepackage{longtable}
\usepackage{graphicx}
\usepackage{newtxtext,newtxmath,amsmath,amssymb}
\usepackage{bm}
\usepackage{color} 
\usepackage{tikz}
\usepackage{units}
\usetikzlibrary{shapes,arrows}
\newcommand{\Cdot}{\!\cdot\!}

\definecolor{Gray}{gray}{0.9}

\begin{document}

\title{Magnetic coupling to the Advanced Virgo payloads and its impact on the low frequency sensitivity}

\author{A. Cirone}
\email{Corresponding Author email address: alessio.cirone@ge.infn.it}
\affiliation{INFN, Sezione di Genova, I-16146 Genova, Italy}
\affiliation{Universit\`a degli Studi di Genova, I-16146 Genova, Italy}
\author{A. Chincarini}
\affiliation{INFN, Sezione di Genova, I-16146 Genova, Italy}
\author{M. Neri}
\affiliation{Universit\`a degli Studi di Genova, I-16146 Genova, Italy}
\author{S. Farinon}
\affiliation{INFN, Sezione di Genova, I-16146 Genova, Italy}
\author{G. Gemme}
\affiliation{INFN, Sezione di Genova, I-16146 Genova, Italy}
\author{I. Fiori}
\affiliation{European Gravitational Observatory (EGO), I-56021 Cascina, Pisa, Italy}
\author{F. Paoletti}
\affiliation{INFN, Sezione di Pisa, I-56127 Pisa, Italy}
\author{E. Majorana}
\affiliation{INFN, Sezione di Roma, I-00185 Roma, Italy}
\author{P. Puppo}
\affiliation{INFN, Sezione di Roma, I-00185 Roma, Italy}
\author{P. Rapagnani}
\affiliation{INFN, Sezione di Roma, I-00185 Roma, Italy}
\affiliation{Universit\`a di Roma "La Sapienza", I-00185 Roma, Italy}
\author{P. Ruggi}
\affiliation{European Gravitational Observatory (EGO), I-56021 Cascina, Pisa, Italy}
\author{B. L. Swinkels}
\affiliation{Nikhef, Science Park 105, 1098 XG Amsterdam, The Netherlands}

\begin{abstract}

We study the electromagnetic coupling of the Advanced Virgo (AdV) Input Mirror Payload (IMP) in response to a slowly time-varying magnetic field. As the problem is not amenable to analytical solution, we employ and validate a finite element (FE) analysis approach. The FE model is built to represent as faithfully as possible the real object and it has been validated by comparison with experimental measurements. The intent is to estimate the induced currents and the magnetic field in the neighbourhood of the payload. The procedure found 21 equivalent electrical configurations that are compatible with the measurements. These have been used to compute the magnetic noise contribution to the total AdV strain noise. At the current stage of development AdV seems to be unaffected by magnetic noise, but we foresee a non-negligible coupling once AdV reaches the design sensitivity.

\end{abstract}

\maketitle

\section{Introduction}

The Advanced Virgo experiment (AdV) \cite{advVirgo,tdr}, hosted by the European Gravitational Observatory (EGO) in Cascina (Pisa), is a Michelson-like laser interferometer endowed with two $3\ km$ long Fabry-Perot resonant cavities and 4 suspended mirror test masses. Its purpose is the detection of Gravitational-Waves (GW) of astrophysical and cosmological origin.

The first detection occurred in September 14, 2015 \cite{bbh1}, when a transient signal produced by the coalescence of two stellar mass black holes was pinpointed from the two advanced Laser Interferometer Gravitational-Wave Observatory (aLIGO) detectors \cite{advLIGO}. From that moment on, several other events were observed \cite{3det,bns,multimess}.

At this stage, AdV can observe a volume of universe 30 times bigger than that accessible to Virgo+, which was the previous detector configuration \cite{advstatus_pub}.
This was made possible by increasing the sensitivity, with the drawback that new noise sources became relevant. In the range of frequency of $10\div100$ Hz, one of the limiting noises could be due to the magnetic coupling through the coil-magnet pairs used as actuators in payloads, which are the mechanical assemblies that suspend the test masses and other ancillary components, including the actuation devices.

This kind of coupling was already observed during the first Virgo Scientific Run (2008), when the substitution of the magnets with five-times smaller ones reduced the magnetic noise contribution to the sensitivity \cite{magnets_pag30}.

Magnetic coupling can also accounts for the correlated magnetic noise from Schumann resonances, which threatens to contaminate the observation of a Stochastic Gravitational-Wave Background (SGWB) in interferometric detectors \cite{thrane2013,schumann_mio}.

In addition magnetic field transients could also enter the analysis pipeline, so that the magnetic coupling can also affect searches for transient GW signals, as reported in \cite{kowalska_2017}.

The first estimation of the magnetic coupling effect had a large uncertainty \cite{tentative} and therefore the overall magnetic noise issue is still open.

In this work we study the magnetic coupling and its impact on the detector sensitivity. We tackle this problem both with Finite Element (FE) simulations and with measurements of the magnetic response of a complex conductive object (the payload), surrounded by a slowly time-varying magnetic field.

Our goal is to determine the magnetic field around a complex, composite object, for a given external magnetic field conditions. For that we need to estimate the detailed electrical configuration and the eddy currents flow, in a situation where we lack a reference standard and the analytical solution is not trivial.

The procedure consists of 3 steps: the construction of the FE model (section \ref{sec:model}), its validation (section \ref{sec:validation}) and the magnetic response estimation in the Virgo environment (section \ref{sec:magnoise}).

\section{The Advanced Virgo payloads}

\begin{figure*}[ht]
\centering
\includegraphics[width=6.69in]{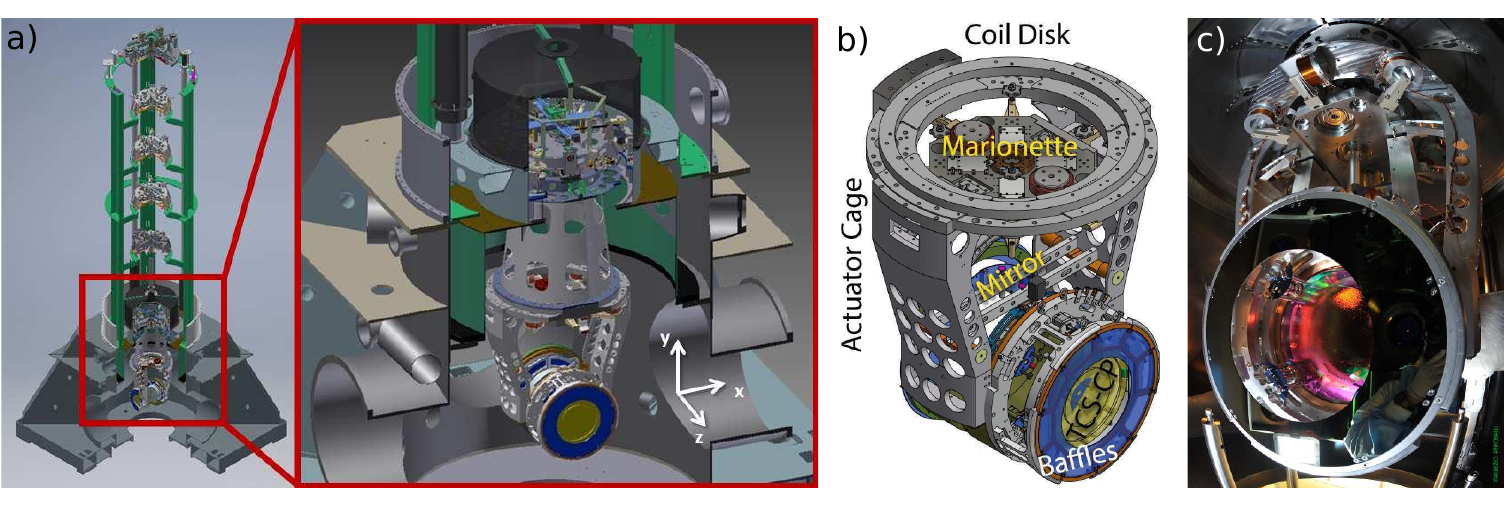}
\caption{a) CAD (Computer-Aided Design) drawings of the AdV Input Mirror Payload (IMP) integrated in the approximately $10$ metres
\textit{Superattenuator} suspension tower; b) CAD drawing of the IMP, including the main metallic assemblies surrounding the TM; c) photo of the integrated IMP.}
\label{fig:Paystruct} 
\end{figure*}

\begin{figure}[ht]
\centering
\includegraphics[width=3.37in]{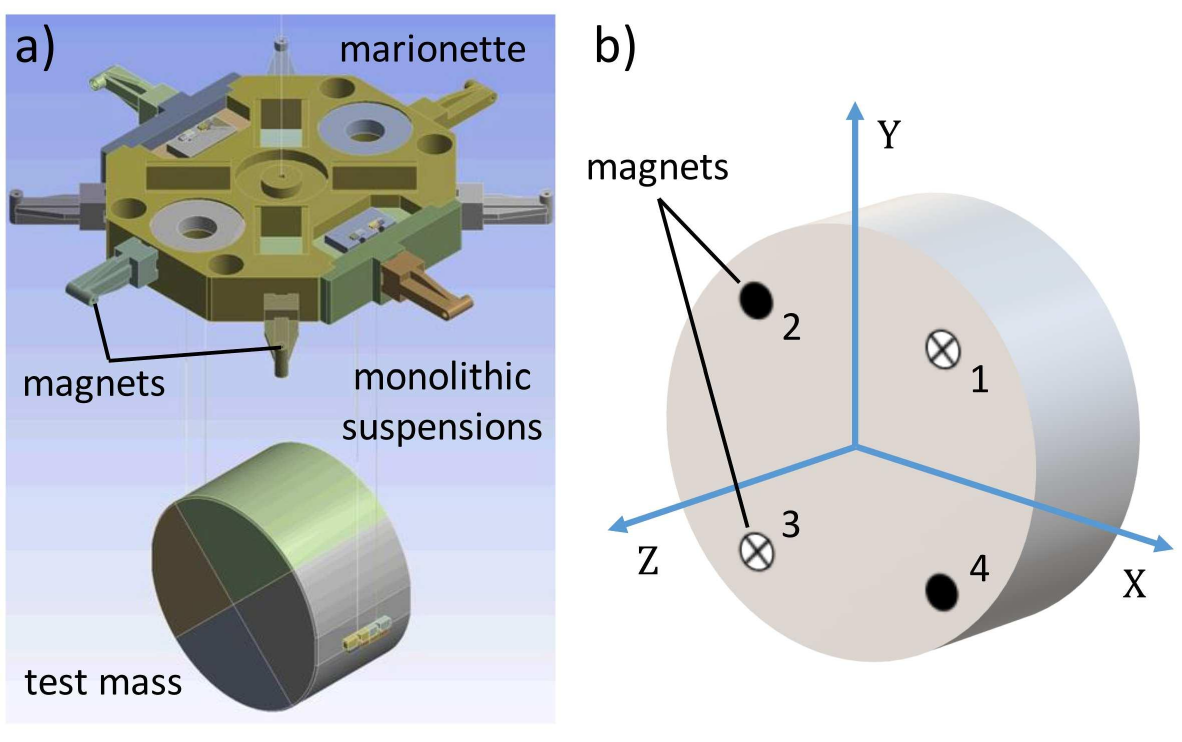}
\caption{a) Schematics of the last suspension stage of the TM, which includes the Marionette with 8 actuation magnets, 4 fused silica suspension fibres and the TM itself; b) schematic view of the cross anti-parallel configuration of the four magnets glued on the TM.}
\label{fig:magnets} 
\end{figure}

The AdV payload (PAY) consists of two suspension stages: the \textit{Marionette}, which is a structure holding the test mass (TM) and the \textit{Actuator Cage} \cite{oldpay_1999}.
The PAY is suspended to the last stage of the so-called Virgo \textit{Superattenuator}, which is a series of six vertical and six horizontal passive mechanical filters (Figure~\ref{fig:Paystruct}a). The overall system is designed to suppress the seismic vibrations by many orders of magnitude, starting from a few Hz \cite{susp}.
The typical configuration of the PAY is shown in Figure~\ref{fig:Paystruct}b.
The main structures close to the magnets are the actuator Cage, the Marionette and a set of ring-shaped components surrounding the TM (\textit{Baffles}, \textit{Ring Heater}, \textit{Compensation Plates - CP}, etc...). 
The Cage is directly connected to the last stage of the Superattenuator through the \textit{Coil Disk}. It also supports the set of driving coils that act on a total of 8 permanent magnets ($Sm_{2}Co_{17}$ magnets of 1 T, 8 mm in diameter and 4 mm thick) glued on the Marionette \cite{tdr}. The magnetic mount of the 8 actuators have horizontal and vertical orientation, in anti-parallel configuration (Figure~\ref{fig:magnets}a). This coil-magnet system steers the PAY in three degrees of freedom: the translation along the beam (roll) and the rotations (pitch and yaw) around the other two orthogonal axes. 
Starting from this general structure, each suspension chain is optimized in different ways so that we have 4 different types of PAY (e.g. the Input Mirror Payload - IMP - in Figure~\ref{fig:Paystruct}c).
Other four coil-magnet pairs act directly on the TM, with the magnets (properties in Table \ref{table_magnets}) glued in a cross anti-parallel configuration (Figure~\ref{fig:magnets}b). This solution is expected to be the most effective against electromagnetic disturbances.

\begin{table}[ht]
\begin{center}
\resizebox{7cm}{!}{
\begin{tabular}{| c | c |}
\hline
\textbf{Properties}  &  \textbf{Values}\\
\hline
Material   &   $Sm_{2}Co_{17}$\\ \hline
Diameter [mm]   &   $1.5 \pm 0.1$\\ \hline
Thickness [mm]   &   $1.5 \pm 0.1$\\ \hline
Residual Induction [T]   &   $1.0 \pm 0.15$\\ \hline
Magnetic moment [mA$\times{m^2}$]   &   $2.0 \pm 0.3 $\\
\hline
\end{tabular}}
\caption{Properties of the magnets glued on the surface of each TM.}
\label{table_magnets}
\end{center}
\end{table}

Indeed these magnets are not only sensitive to the magnetic field produced by the driving coils but also to any (noisy) external magnetic gradient. While the anti-parallel configuration of all the magnets located on both Marionette and TM should be insensitive to any contribution that is spatially uniform, asymmetries due to a distinctive field-structure can produce a net total force, causing a displacement noise. In addition the magnetic forces act on the TM directly, by-passing all the seismic isolation provided by the suspension system.

\section{Magnetic coupling to the payloads}
\label{sec:magcoupling}

Magnetic disturbances can be expressed as a superposition of different magnetic contributions derived both from known point-like sources (local) and unknown ones (environmental).
In the estimation of the magnetic noise, we considered only environmental components because it is assumed that there are no nearby magnetic sources. 
Under this hypothesis, the contribution of this noise to the AdV noise budget is relevant only in the low frequency range because at high frequencies (roughly above $150$ Hz) the interferometer sensitivity is dominated by other contributions while the environmental magnetic field (at the PAY location) is filtered out by the metallic enclosure which surrounds the PAY.
In the noise budget estimation we explicitly consider the low-pass filtering effect of the steel tank. 

A magnetic field acts on a permanent magnet causing the following two effects: $\emph{(i)}$ translational force associated to the magnetic gradient: $\textbf{F}=\nabla(\bm{\mu}\cdot \textbf{B})$, were $\bm{\mu}$ is the magnet's magnetic moment; $\emph{(ii)}$ torque produced by the magnetic force: $\tau=\textbf{F}\times\textbf{r}$.
Actually we should talk about two distinct components of torque: one related to the TM and the other related to each single magnet. However, in point-like magnet approximation we neglect the second contribution, especially because the magnets are glued on the TM and the whole system is treated as a rigid body.

In Advanced Virgo, all the magnets used for the TMs actuation have approximately the same magnetic moment, directed along the beam propagation direction (i.e. $\bm{z}$). 
The force on each magnet can be written as $F_{z}=\mu(\partial B/\partial z)$, so the total force on the TM is simply the sum of the forces on the four magnets.
In the ideal situation in which the four magnets have exactly the same magnetic moment and if the magnetic field gradient is the same on each magnet, the anti-parallel cross configuration should guarantee a null force on the TM.
In the real-world case, $\bm{\mu}$ has a tolerance of about $15\%$ around the nominal value and the magnetic field gradient spatial symmetry is not guaranteed.

The main cause for a non-uniform gradient is the interaction between the external field and the metallic structure of the PAY. This structure couples with any time-varying magnetic field generating eddy currents, which in turn warp the field and produce a gradient.

We study only the IMP, because -- together with the End Mirror Payload (EMP) -- it is the most sensitive to magnetic coupling.

The study of the magnetic response of an object is a classical electromagnetic problem where analytical solutions exist only for very simple geometries.
If the goal is to know the magnetic field in each point of a volume, direct measurements are also impractical. 
The typical approach is to use a numerical solution (i.e. Finite Element (FE) analysis).

\section{System modelling}
\label{sec:model}

The PAYs are very complex objects consisting of several parts (ten main parts; volume of about $1.2$ $m^3$) of different materials that are assembled by bolts, welds or screws.
We chose to use the COMSOL Multiphysics FE analysis simulation software \cite{fe_book} and the AC-DC module for the computation of our time-varying magnetic field studies. For a closer look to some delicate simulation steps (e.g. geometry simplification and model meshing), see the supplementary material (section I).

A subtle but significant problem lies in how parts are assembled. When simulating two adjacent metallic domains (i.e. in close physical contact), the software implicitly assumes electrical conductivity between them. 
In practice though, when two metallic parts are assembled without welding, the electrical contact can be impaired by a thin insulating film (e.g. oxides, contaminants, reaction products), which introduces an energy barrier that can limit the current flow \cite{holm,elcontact}.
Depending on the thickness of the barrier, electrons may not have enough energy to tunnel across it. In particular, eddy currents are very feeble and therefore they are especially susceptible to energy barriers. 
The conductivity across two surfaces in electric contact is determined by several factors: the type of material, the surface finishing and the pressure applied between the two surfaces. 
We designed and validated a simple model (open coil) which verified that the eddy currents flow critically depends on the force applied to keep two aluminium surfaces in contact. We experimentally measured a sharp transition of the eddy currents with respect to the force applied.
This simple experiment prompted us to consider each connection among the metallic parts of a composite object as a two-state system (open-closed) in order to take into account the (unknown) state of the surfaces and the applied pressure.

The subdivision of a complex object such as the PAY into disjointed parts can be parametrized by the conductivity at the level of each mechanical connection $p$ as 1 or 0, leading to a very large number of configurations ($2^p$). 
Therefore we employed a hierarchical analysis to capture only the most relevant contact points in the PAY assembly.

\begin{figure}[ht]
\centering
\includegraphics[width=3.37in]{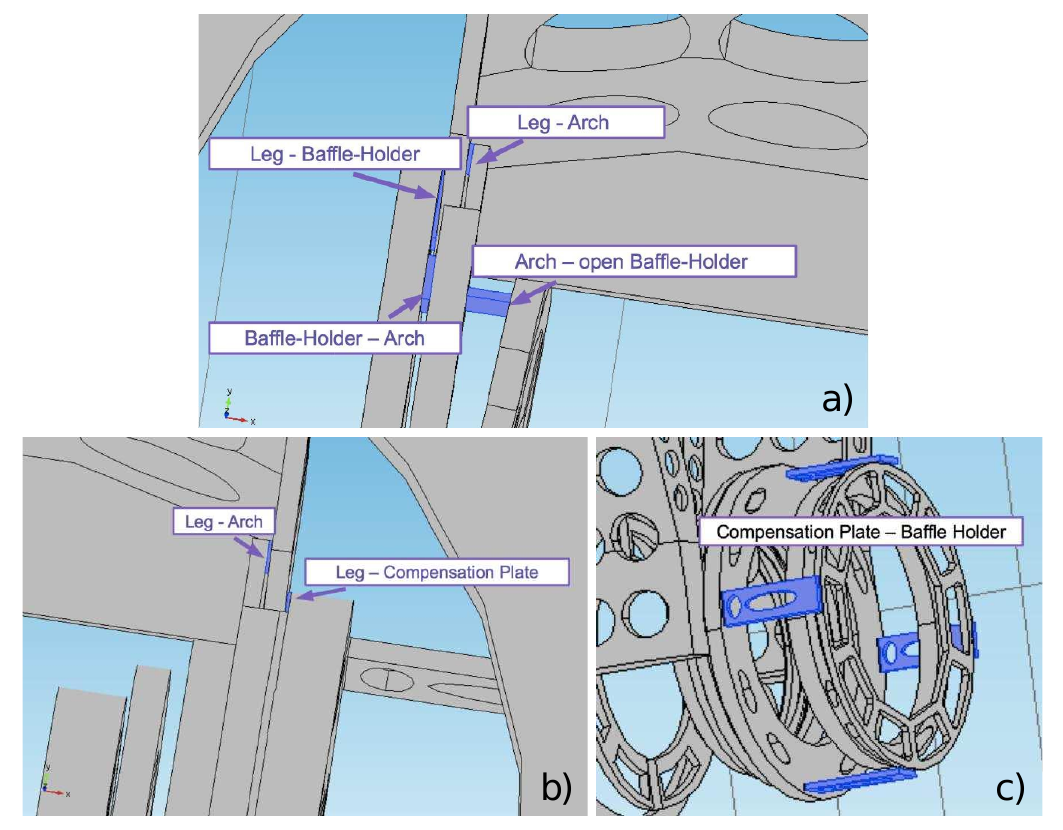}
\caption{Mechanical connections used to simulate two-state electrical system: a) \textit{Cage Legs-Baffle Holder}, \textit{Cage Legs-Arch}, \textit{Arch-Baffle Holder} and \textit{Arch-open Baffle Holder}; b) \textit{Cage Legs-Compensation Plate} and \textit{Cage Legs-Arch}; c) \textit{Compensation Plate-Baffle Holder}.}
\label{fig:7conn} 
\end{figure}

Overall, we found $p=7$ main connections which have a relevant impact on the gradient nearby the magnets. The corresponding parametrized volumes were inserted in correspondence of the real mechanical connections (screws and bolts), as presented in Figure~\ref{fig:7conn} and they were modelled in COMSOL using either air (insulation) or aluminium (same material as the bulk).

The PAY assembly is made of different materials: aluminium, steel and titanium. In the FE analysis we modelled only the aluminium components, as they are the most relevant ones in terms of number of parts and they have the highest conductivity, $3.03\Cdot10^{7}$ $S/m$, compared to $1.4\cdot10^{6}$ $S/m$ for steel and $0.6\cdot10^{6}$ $S/m$ for titanium.

The model was embedded in a uniform, sinusoidal magnetic field of frequency $f=\omega/2\pi$.

\section{Model validation}
\label{sec:validation}

\begin{figure*}[ht]
\centering
\includegraphics[width=5in]{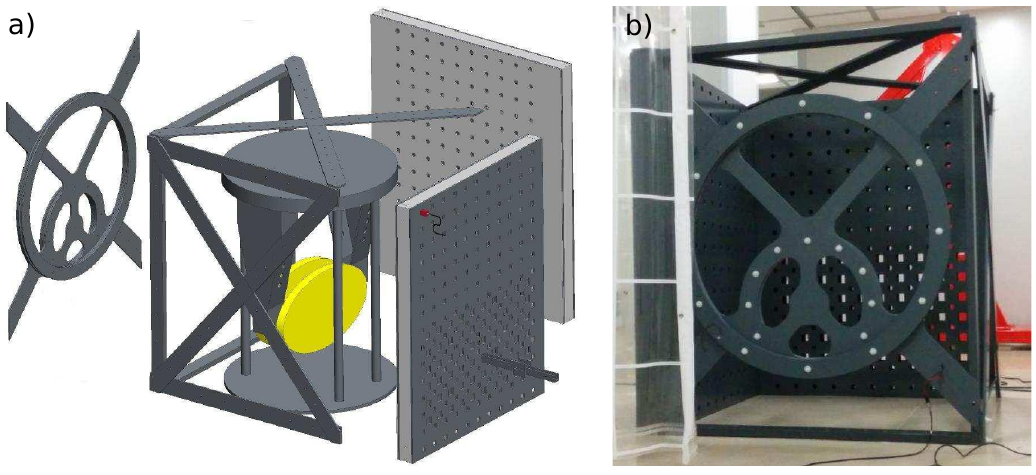}
\caption{The PVC frame built for the FEA validation: CAD drawing (a) and the actual object inside the Virgo clean room (b). The structure embeds two integrated coils on one side and two gridded panels on the opposite sides for accurate positioning of the magnetic probe.}
\label{fig:cage} 
\end{figure*}

The FE model delivers the electromagnetic field at any point but we still have to determine the electrical connection states among the relevant parts of the assembly (model validation).
This means that we have to find out which of the $2^{7}=128$ possible models represents the real PAY at best, that is what we need to compare each FE simulation with an experimental measure.
In addition, we can use the measures to tune the geometrical representation and the uncertainties on the material properties.

The validation procedure requires an accurate setup, where the driving magnetic field and the geometrical properties are well under control, so that discrepancies between the simulation and the measurements can be assigned to the model assumptions. 

\subsection{Experimental set-up} 

We built a driving system with two coils ("Big" and "Small", with an external radius of respectively $1.040\pm 0.001\ m$ and $0.540\pm 0.001\ m$) and a structural PVC cubic frame that aligns them with the PAY (Figure~\ref{fig:cage}).
The coils are made of 100 turns of copper wire with a diameter of $0.95\pm 0.02\ mm$. 
The PAY was housed between the coils and two gridded panels with several reading slots, where we can insert a magnetic probe: this is a triaxial magnetic field sensor FL3-100 (Stefan Mayer Instruments), with intrinsic noise $<20\ pT/\sqrt{Hz}$ at $1\ Hz$ and a measurement range of $\pm 100\ \mu T$. 
The slots are geometrically tailored to the probe in order to minimize positional errors. They are designed to get a fairly dense sampling of the magnetic field. 
Some of the holes are bigger so that a probe-holder rod can be inserted, in order to make spot measurements along a line inside the volume of the PAY. A picture is provided as supplementary material (Figure S1). 
The two coils generate spatially different field configurations and therefore provide a more accurate validation. They can be independently driven by an AC current generator (CoCo80, Crystal Instruments) coupled to a linear amplifier (BAA 120, TIRA).
The cubic frame was placed in the Virgo Central Building (CEB), in a class $100$ clean room under the input towers. We ensured that there were no significant metallic objects in a radius of $\approx 2\ m$ around the apparatus (Figure S2 of the supplementary material).

\subsection{Reference measurements} 

We began our validation with the "zero" measure, that is a set of reference measurements to study the contribution of the surrounding environment and the cubic frame alignment. 

The measurements were done using the experimental setup without the PAY inside it. 
Occasionally the amplifier exhibited a small drift from the nominal value of $1$ A, which was taken into account in the post processing analysis by performing a current optimization. 
The three components of the magnetic field were measured on a set of $65$ holes for each panel.

\begin{figure}[ht]
\centering
\includegraphics[width=3.37in]{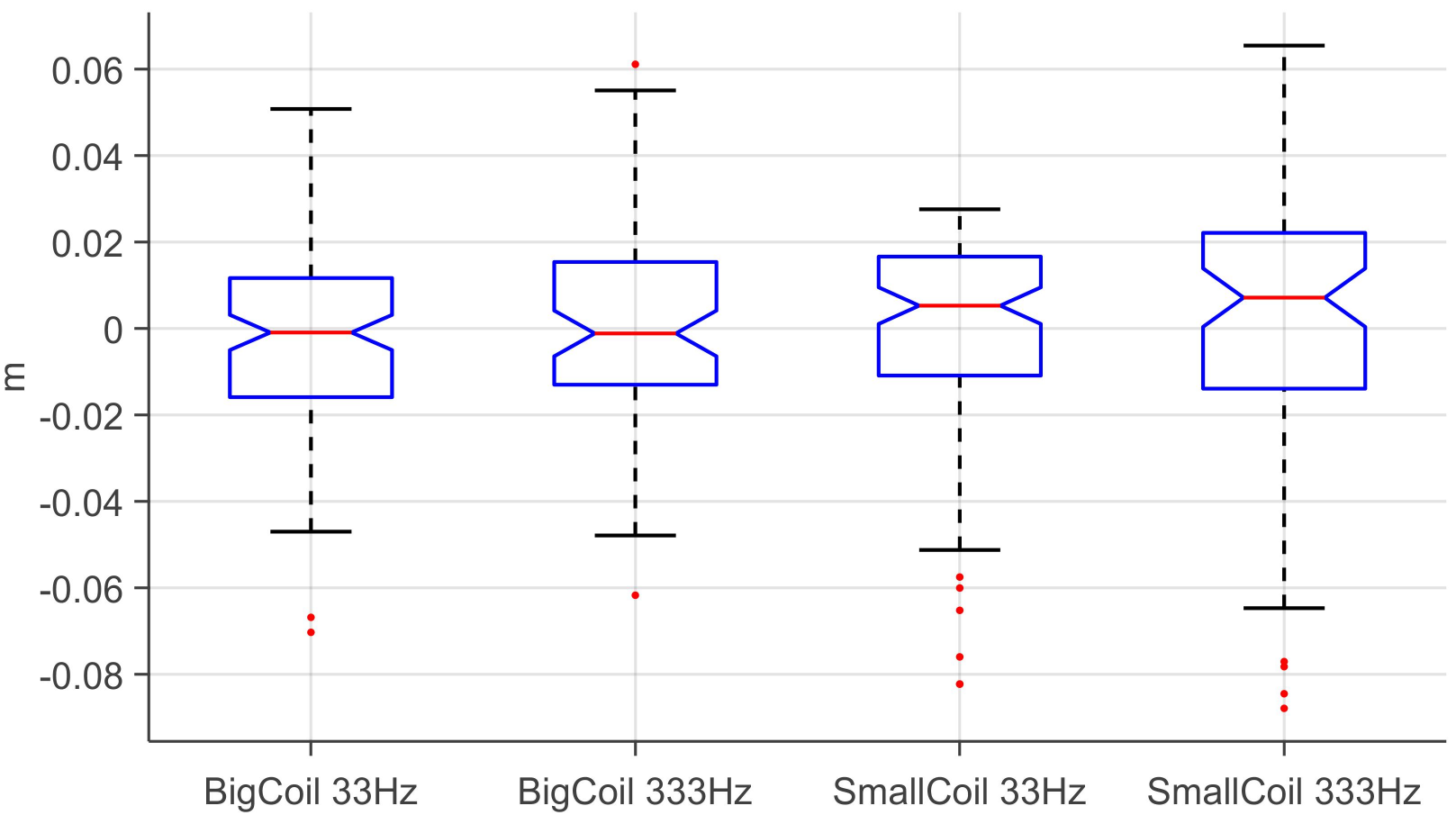}
\caption{Distributions of $m$ without PAY (Eq. \ref{eq:comparison}, reference measure). The statistics is computed on all the measured positions, for both coils and two frequencies. The red horizontal line is the median value while the bottom and top edges of the blue boxes indicate the 25th and 75th percentiles, respectively. The whiskers extend up to 1.5 of the box range and the outliers are plotted individually using single red dots.}
\label{fig:ConfigRelErr} 
\end{figure}

At the same time we simulated the structural frame and the coils. The PVC frame is transparent to magnetic fields, allowing us to consider only two materials in the simulation: copper for the coil windings and "air" for the remaining parts. 
The magnetic field was calculated in correspondence of the measurement points in all the explored configurations. 
Finally we compared the experimental and simulated data by minimizing the relative difference function $m_{k}$, with $k=[1,\ $number of measuring points$]$, between the measured and the simulated magnetic field value over the input current:

\begin{equation}
m_{(k)}=\left|\frac{B_{meas}^{(k)}-cB_{sim}^{(k)}}{B_{meas}^{(k)}}\right|
\label{eq:comparison}
\end{equation}
where $c$ is the optimization parameter.
Measurements and simulation were found to agree within $\approx 5\%$, after the current optimization. Looking at the data (Figure~\ref{fig:ConfigRelErr}), we assess that the distribution of the relative differences is compatible with zero for both the "Big" and "Small" Coil configurations and at both $33\ Hz$ and $333\ Hz$. 

\subsection{Input Mirror Payload measurement} 

\begin{figure*}[htb]
\centering
\includegraphics[width=5in]{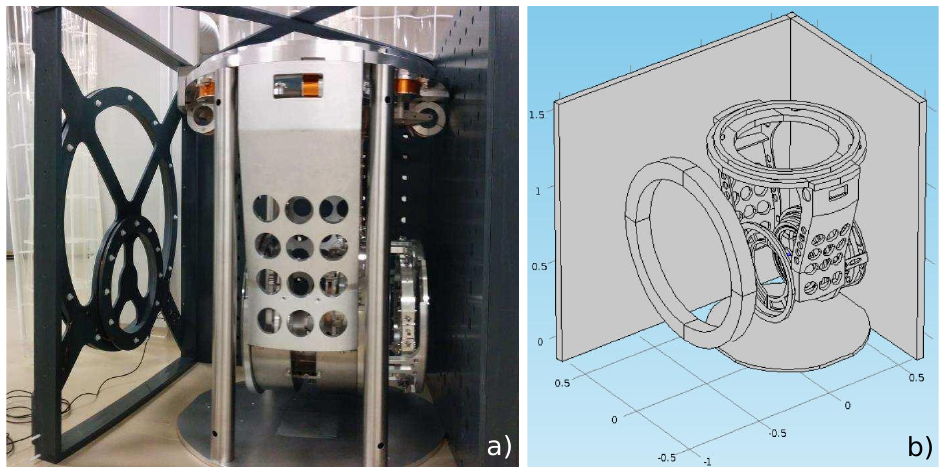}
\caption{The IMP inside the PVC frame: photo of the experimental apparatus (a) and the FE model (b).}
\label{fig:IMPmeas} 
\end{figure*}

The entire procedure was repeated with the IMP inside the frame. The PAY was kept in place by a support structure with an aluminium base and four steel legs (Figure~\ref{fig:IMPmeas}a).
The PAY was laser-aligned with the coil axis (estimated accuracy of $\simeq 1\ cm$) and measurements were taken both in correspondence of the two gridded panels and with the help of the extension rod, to get as close as possible to the PAY assembly.

The validation goal is to identify which electrical configuration best represents the real PAY. The procedure is detailed in the supplementary material (section II).

\begin{table}[b]
\begin{center}
\resizebox{8cm}{!}{
\begin{tabular}{| c | c |}
\hline
\textbf{}  &  \textbf{$(B_{meas}-B_{sim})/B_{meas}$}\\
\hline
payload - "Small" Coil   &   $-0.031 \pm 0.021$\\ \hline
payload - "Big" Coil   &   $-0.0057 \pm 0.0220$\\ \hline
reference - "Small" Coil   &   $0.0052 \pm 0.0090$\\ \hline
reference - "Big" Coil   &   $-0.00093 \pm 0.00740$\\ \hline
\end{tabular}}
\caption{\textit{Mean and standard deviation of $m$ (Eq. \ref{eq:comparison}) with and without the PAY structure. Values combine both frequencies (}$73$ Hz\textit{ and } $33$ Hz\textit{) and are given for the two source configurations ("Big" and "Small" Coils).}}
\label{agreement}
\end{center}
\end{table}

Eventually, we identified 21 statistically equivalent configurations that can represent the real object: the equivalence was defined by using the non-parametric Kolmogorov-Smirnov test, applied to the sorted distribution versus the best one.
Since all these configurations are statistically equivalent, all of them have to be considered in the evaluation of the magnetic noise contribution to the Advanced Virgo sensitivity.
The agreement between measurements and simulations is presented in Table~\ref{agreement}, where also the reference case at 33 Hz is reported once more for comparison. These values are the overall statistics of the 21 equivalent configurations.

\section{Magnetic strain noise}
\label{sec:magnoise}

Within a frequency domain representation, a force $F$ acting on one interferometer TM displaces it, along the laser beam propagation direction (namely \textbf{z}), by $\Delta L = F/(M\omega^2)$, where $M$ is the TM mass (kg) and $\omega = 2\pi f$ is the frequency in $rad/s$. The associated strain noise in the detector is:

\begin{equation}
h_{mag}=\frac{\Delta L}{L_{0}}
\end{equation}
where $L_{0}=3000$ m is the interferometer arm length.

The total $h$ due to the PAY magnetic coupling is the sum of a translational term with a rotational one. Assuming linear superposition and assuming that the magnetic forces acting on the TM are uncorrelated over long range -- so that the sum over the 4 TM is treated as incoherent -- the term associated to the translational force is:

\begin{equation}
h_{transl}=\frac{1}{L_{0}M(2\pi f)^{2}}\sqrt{\sum_{j=1}^{4}\big({\sum_{i=1}^{4}F_{i}}\big)_{j}^{2}}
\label{htr1}
\end{equation}
with $M=42$ kg the mass of the AdV TM \cite{tdr} and $f$ the frequency. For each $j$-th TM, the total magnetic force is the sum of the force on the $i$-th magnet.
Considering that all the magnetic moments are directed along the \textbf{z} direction and that they are constant, the translational magnetic contribution is reduced to:

\begin{equation}
h_{transl}=\frac{1}{L_{0}M(2\pi f)^{2}}\sqrt{\sum_{j=1}^{4}\big[\sum_{i=1}^{4}\mu_{i}\big(\frac{\partial B_{z}}{\partial z}\big)_{i}\big]_{j}^{2}}
\label{htr2}
\end{equation}
where $B_z$ is the magnetic field component along \textbf{z}. As the force is directed along \textbf{z}, only two kind of torques exist: $\tau_x$ and $\tau_y$. Considering $F^{j}_{z}$, with $j=[1,4]$, the magnetic forces on the four magnets (as defined in Figure~\ref{fig:magnets}b), we have:

\begin{equation}
\tau_{x}=[(F_{z}^{1}+F_{z}^{2})-(F_{z}^{3}+F_{z}^{4})]y
\end{equation}
\begin{equation}
\tau_{y}=[(F_{z}^{1}+F_{z}^{4})-(F_{z}^{2}+F_{z}^{3})]x
\end{equation}
where $x$ and $y$ are the components of the magnet position vector or, in other words, the force's application point relative to the TM centre mass. Hence the rotational contribution to the strain is:

\begin{equation}
h_{rot}=\frac{D}{L_{0}(2\pi f)^{2}}\sqrt{\sum_{j=1}^{4}\big(\frac{\tau_{xj}}{I_{xx}}+\frac{\tau_{yj}}{I_{yy}}\big)^{2}}
\end{equation}
where $I_{xx}$ and $I_{yy}$ are the moments of inertia of the TM and $D$ is the laser misalignment from the centre of mass of the TM (assumed the same for all TMs). Finally we can compute the total magnetic strain noise, considering the same contribution on each TM:

\begin{equation}
h_{mag}=h_{transl}+h_{rot}=\frac{2F}{ML_{0}\omega^{2}}+\frac{2\sqrt{2}D}{IL_{0}\omega^{2}}(\tau_{x}+\tau_{y})
\label{hmagn}
\end{equation}
with $F$, $\tau_{x}$ and $\tau_{y}$ respectively the total force and torques calculated on each TM. We also assumed identical inertia moments ($I_{xx}=I_{yy}=I$) and a conservative beam off-centering of $D=1$ mm.

\subsection{Environmental magnetic field measurement}
\label{subsec:benv}

\begin{figure}[ht]
\centering
\includegraphics[width=3.37in]{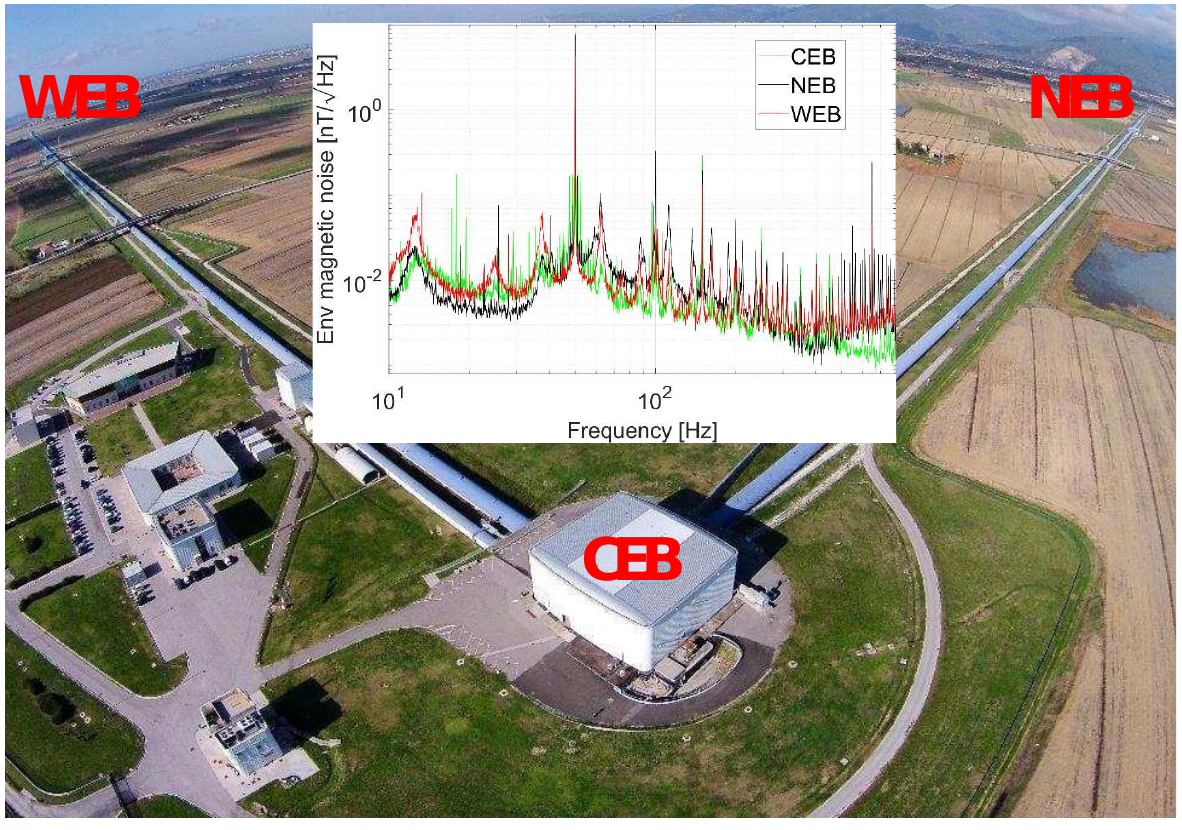}
\caption{Aerial photo of Virgo site. In the inset: the environmental magnetic field activity spectrum, acquired in the three main buildings (CEB = central, WEB = west end, NEB = north end).}
\label{fig:benv} 
\end{figure}

We based our estimation on a set of measurements carried out in August 2017. We sampled the magnetic field at several locations inside the three experimental Virgo buildings: the Central Building (CEB), where the two input TMs are located, and the two end buildings (North End Building - NEB; West End Building - WEB), which host the two end TMs.
For each building, we computed the magnitude of the magnetic field vector (Figure~\ref{fig:benv}).
Then we took into account the filtering effect produced by the steel tank that surrounds the suspension system, which effectively acts as a first order low pass filter with a cut-off frequency of $f_{0}=5$ Hz. 
Thus our estimate of the equivalent $B_{env}$ at the TMs is

\begin{equation}
B_{env}=\frac{\sqrt{B_{CEB}^{2}+B_{CEB}^{2}+B_{NEB}^{2}+B_{WEB}^{2}}}{\sqrt{1+(\nicefrac{f}{f_{0}})^{2}}}
\label{Benv}
\end{equation}

\subsection{Force calculation}

The FE model of the PAY was placed in a uniform magnetic field $B=1$ T directed along the \textbf{x}, \textbf{y} and \textbf{z} axis and with frequency range $f=[10, 2000]$ Hz. We observed that the main contribution to the gradient came from $B_{z}$.
The force on each magnet is computed according to section \ref{sec:magcoupling} and the net total force is the mean value of a Monte Carlo procedure to take into account deviation from the nominal parameters both in the magnet moments and in the magnet positioning (details in supplementary materials, sections III-V).
This procedure was iterated for each of the 21 electrical configurations.

\section{Results}

\begin{figure}[ht]
\centering
\includegraphics[width=3.37in]{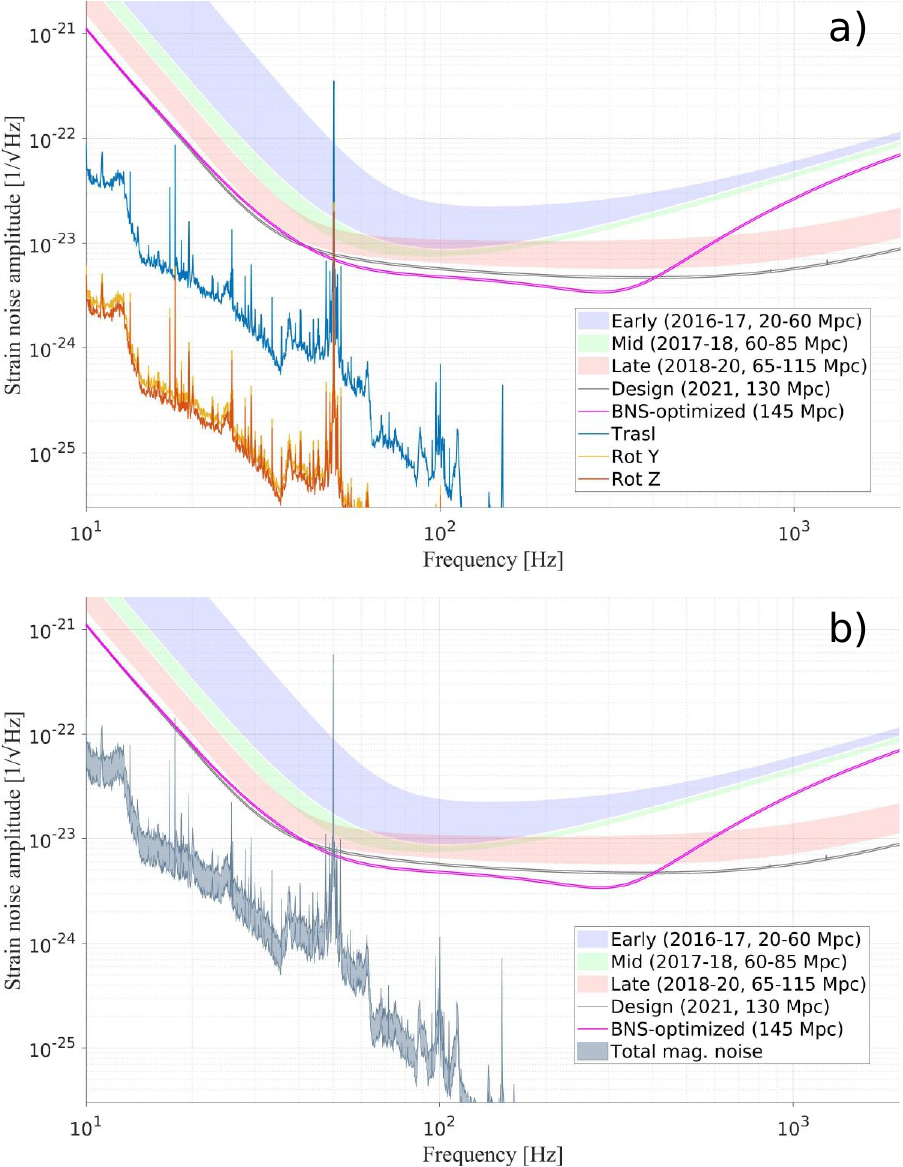}
\caption{Contribution of the magnetic noise to the AdV sensitivity, in the frequency range of interest for astrophysical binary mergers detection. a) Spectra of both translational (blue line) and rotational (red and orange lines) magnetic effect on the TMs; b) Total spectra envelope (translational plus rotational) on all 21 electrical configurations.}
\label{fig:spec} 
\end{figure}

In Figure~\ref{fig:spec}a we show the estimation of the magnetic contribution to the AdV strain noise, calculated by multiplying $h_{magn}$ in equation \eqref{hmagn} by the real magnetic field spectrum of equation \eqref{Benv}. The force values used in this plot are an average on the 21 equivalent electrical configurations.  

The translational force contribution (blue line) is dominant on both rotational ones (red and orange lines). The comparison with the projected sensitivity curves shows that magnetic noise (translational plus rotational) could be considered a nuisance only for the latest interferometer specifications (design -- gray curve). In general though, the magnetic contribution stays well below the safety requirements (one order of magnitude below the best sensitivity limit).

When we consider the uncertainty due to the electrical configurations, we see that the spread in the calculated force values is well within safe limits. Figure~\ref{fig:spec}b  shows the envelope region due to slightly different forces values computed on the 21 electrical configurations. In this plot we also consider the combined contribution of both translational and torque forces. 

Naturally, the upper expected noise gets closer to the sensitivity curves although is still of moderate concern for the latest design specification only.

\section{Discussion}

The validation step had the goal to define the electric configuration of the PAY and the agreement between the simulated and the measured field was found to be within $5\%$. 
This value is the current limit of the FE model in representing the real PAY and it is due to all the uncertainties and simplifications embodied both in simulations and in the experimental step.

The hierarchical and simplification steps are a reasonable heuristic approach to a problem with this complexity. Yet, these steps could have introduced further uncertainties, hence the need for further refinement and study of different decompositions.
Moreover, the steel and titanium parts of the PAY were neglected in the simulation, given that their contribution is suppressed by a factor $\approx 20-30$ due to the electrical conductivity.

Forces and displacements were calculated supposing a uniform environmental magnetic field in the area around the PAY. This assumption is almost surely not verified in the real conditions.  
The only practical way to test the simulation forecasts is to measure the transfer function during AdV working conditions (i.e. during the commissioning phase).

The displacement produced by the force on the TMs summarizes all the forces acting on the four TMs of the two Fabry-Perot cavities. We assumed identical PAY structures and uncorrelated contributions, so that the total displacement is calculated in quadrature.
These assumptions stem from the close similarity among the IMP and the EMP (the main difference is in the compensation plate, which is present only in the IMP) and the significant distance between them. That said, the TMs in the central building are rather close so that correlation in the magnetic response cannot be excluded. This issue is not considered here and needs further testing. Moreover we are going to address the Schumann resonances common-mode contribution in a separate study, which lays its foundation on the paper \cite{schumann_mio}.

Results show that the magnetic noise budget should not impact the AdV initial observations (2017-2019). Nevertheless, in a couple of small frequency ranges, the estimated noise is higher than the desired level (technical noises are required to be $\leq 0.1 $ of the incoherent sum of all fundamental noises).
If these simulations are confirmed by experimental data on the interferometer, this would prompt the drafting of mitigation strategies.
For instance, the environmental field was already addressed by past mitigation actions carried out on initial Virgo (2009). These included the size reduction of TM magnets and power cable routing optimization.

As a direct consequence of this work, the design of the PAY structure was optimized to reduce gradients and shorten eddy currents paths.

The agreement between simulation and measurement underlines the importance of a extensive validation phase; this would imply a more realistic modelling of the PAY environment and measurements on the interferometer with dedicated magnetic injections. We would therefore have a detailed comprehension of local and environmental magnetic interaction which could be used to plan more effective mitigation strategies, if needed.

\section{Conclusions}

The sensitivity of the upgraded gravitational interferometer Advanced Virgo is limited by different kinds of noises. In the low frequency range, we focused on the noise associated to the coupling between the environmental magnetic field and the PAYs. Any time-varying magnetic field interacts with the metallic structure of the PAY to produce local gradients, which exert forces on the four magnets glued on the TM. These can induce a worsening of the intrinsic displacement noise of the detector and, in the presence of high magnetic transients, create a glitch, which can be misinterpreted as a GW signal. 

In order to understand the contribution of the magnetic noise, we studied the magnetic response of the PAY to a given environmental field. 
Since a direct measurement of the induced field is impractical, we had to use FE simulations. 

Several steps were performed on the system in order to reduce the solution time and simplify simulation. Then, an optimization procedure based on Design of Experiment (DoE) techniques was developed to find the optimal electric configuration of the PAY. 

The magnetic field gradient was calculated taking into account magnetic moment and magnet position dis-homogeneities.
Monte Carlo simulations were used to compute the total force on the TM as a function of the magnetic moment tolerance and of the magnet position error, with respect to the PAY structure. 

The strain noise contribution was estimated for both translational and rotational forces and it was compared with the AdV sensitivity curve. 
Results suggest that the magnetic noise contribution to the strain is not dramatic for the time being, while it will possibly be an issue when the detector will approach its final stage design sensitivity and beyond it. 
For this reason we are already working on performing further measurements to refine the analysis and compare it to measurements in the actual AdV working conditions. 

\section*{Supplementary material}

See supplementary material for further details on the system modelling, the experimental set-up and the magnetic force computation.

\section*{Acknowledgements}

The authors gratefully acknowledge the European Gravitational Observatory (EGO) and the Virgo Collaboration for providing access to the Central Building clean room and to the environmental data. AC and MN was supported by the INFN Doctoral Fellowship at the University of Genova.

\bibliographystyle{unsrt}
\bibliography{Mendeley}

\end{document}